\documentclass[]{bytedance_seed}

\usepackage{amsmath,amssymb,mathtools}
\usepackage{array}
\usepackage{float}
\usepackage[toc,page,header]{appendix}

\newcommand{\modelname}{HELIX}
\providecommand{\Description}[1]{}
\hypersetup{
  pdftitle={HELIX: Purified and Unified - Rethinking Feature Interaction and Sequence Modeling for Large-Scale Recommendation},
  pdfauthor={Data - Global E-Commerce Recommendation Video Team}
}

\title{\modelname: Purified and Unified\,\textemdash\,Rethinking Feature Interaction and
Sequence Modeling for Large-Scale Recommendation}

\author{Data - Global E-Commerce Recommendation Video Team}

\abstract{
Industrial recommendation ranking models typically scale along two modeling
axes: feature interaction over heterogeneous user, item, context, and cross
features, and sequence modeling over long, informative, and multi-type user
behavior histories.
We find that scaling either capability in isolation is insufficient, as each
exhibits a limited scaling ceiling and a suboptimal scaling-law slope. We
conjecture that achieving a more favorable scaling-law slope requires jointly
scaling both axes. To support this, we present \modelname, a purified and unified
architecture for large-scale recommendation. \modelname{} interleaves sequence
retrieval and feature interaction while enforcing one-way information flow from
reusable sequence states to candidate-conditioned mix-tokens.
This design preserves cross-depth communication between the two modeling axes
while keeping user-side sequence computation amortizable, enabling flexible and
asymmetric scaling of sequence modeling and feature interaction. Deployed in
TikTok's e-commerce recommendation system, \modelname{} consistently improves
offline CTR AUC, CVR AUC, and other ranking metrics. In online A/B tests, it achieves an
approximately 6\% increase in e-commerce video GMV per user.
}

\date{August 27, 2026}

\begin{document}
\maketitle

\vspace{-6mm}
\begin{figure}[H]
  \centering
  \includegraphics[width=1.0\textwidth]{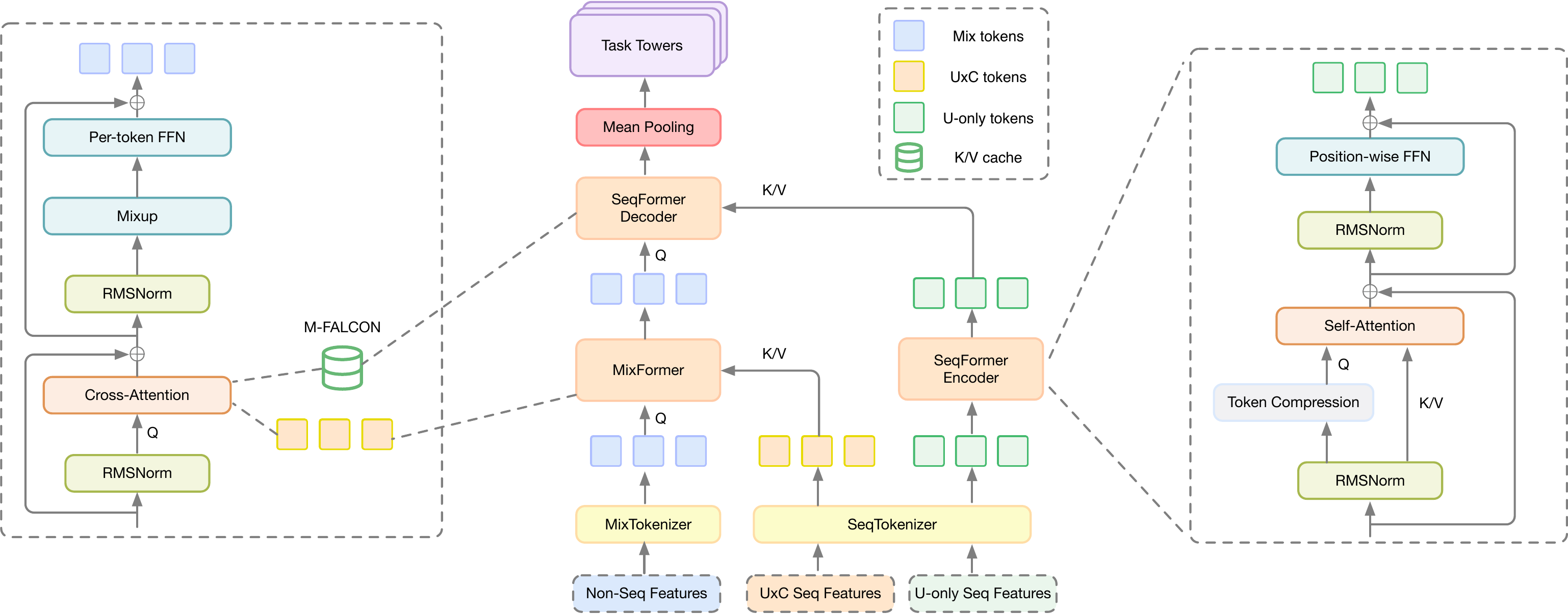}
  \caption{Architecture of \modelname. Non-sequence, U$\times$C sequence, and U-only
  sequence features are tokenized into separate streams. The model stacks three
  types of blocks: (1) SeqFormer Encoder blocks encode U-only sequences into a
  reusable K/V cache; (2) MixFormer blocks repeatedly retrieve from U$\times$C sequences
  and update mix-tokens through feature interaction; and (3) SeqFormer Decoder
  blocks repeatedly retrieve from the cache and further refine mix-tokens through
  feature interaction.}
  \Description{The center shows three input streams. Non-sequence features are
  tokenized into mix-tokens, while U$\times$C and U-only sequence features are
  tokenized separately. U-only tokens pass through a SeqFormer encoder, whose
  final retained states are projected into a reusable user-side K/V cache.
  MixFormer combines mix and U$\times$C tokens and retrieves sequence information
  before the SeqFormer decoder, mean pooling, and task towers produce outputs.}
  \label{fig:architecture}
\end{figure}

\clearpage
\setcounter{tocdepth}{2}
\tableofcontents
\clearpage

\section{Introduction}
\label{sec:introduction}

Industrial recommendation ranking models sit at the junction of two modeling
problems. The first is \bodyemph{feature interaction}: given many user, item,
context, and cross features, the model must discover high-order combinations
that are predictive of classification and regression targets such as CTR, CVR,
and GMV.
Classical architectures such as FM, DeepFM, and DCN
\cite{rendle2010fm,guo2017deepfm,wang2017dcn} encode manually designed
interaction patterns through strong inductive biases. They perform well at moderate
model scales, but their gains often saturate quickly as width and depth
increase, and larger variants can become increasingly difficult to optimize. The second is \bodyemph{sequence modeling}: the model must interpret
behavior trajectories to identify user intent and candidate-relevant evidence.
Target-aware sequence models such as DIN, DIEN, and DSIN
\cite{zhou2018din,zhou2019dien,feng2019dsin} have made candidate-conditioned
behavior aggregation a standard component in industrial ranking systems, but their
relatively shallow aggregation limits gains from increased
sequence-modeling capacity.

Recent work has begun to unlock scaling along each axis independently. For
feature interaction, Wukong, RankMixer, and TokenMixer-Large
\cite{zhang2024wukong,zhu2025rankmixer,jiang2026tokenmixer} 
tokenize heterogeneous feature embeddings and apply stable residual blocks,
making feature interaction resemble a scalable token-processing backbone.
On the sequence-modeling side, HSTU, Ultra HSTU, and STCA
\cite{zhai2024actions,ding2026bending,guan2026make} show that richer attention
blocks can produce predictable gains as sequence modeling capacity grows.

However, scaling either axis alone leaves the other as a bottleneck. Recent
unified architectures such as InterFormer \cite{zeng2025interformer}, Kunlun
\cite{hou2026kunlun}, and OneTrans \cite{zhang2026onetrans} bring feature
interaction and sequence modeling into a shared backbone, enabling the two
capabilities to communicate across stacked layers and scale jointly.

Unified designs face three limitations. First, TokenMixer and UniMixer
\cite{jiang2026tokenmixer,ha2026unimixer} suggest that
self-attention is not necessarily the most effective feature-interaction
operator for heterogeneous tokens. Second, one-step designs such as OneTrans, SORT \cite{wang2026sort},
and MTFM \cite{song2026mtfm} perform feature interaction and sequence retrieval
within the same modeling step, which may disperse mix-token attention over
heterogeneous non-sequential inputs and weaken behavior retrieval. Third, bidirectional
designs such as InterFormer and Kunlun allow candidate-conditioned features to
update user-side sequence states, making those states candidate-dependent and
preventing user-side computation from being amortized across candidates with
M-FALCON \cite{zhai2024actions} or across samples with user-level training in
production settings.

To address these limitations, we present \modelname, a purified and unified architecture for
large-scale recommendation. It is \bodyemph{unified} because sequence retrieval and
feature interaction are interleaved through the network rather than separated
into two stages. It is \bodyemph{purified} because it avoids extra architectural
components and training objectives, such as LHUC, DCN-M, FM modules, cross-layer
residual connections, and auxiliary losses, retaining only the essential
operations for feature interaction and sequence modeling.
The backbone also enforces a one-way information flow from the reusable
sequence-modeling branch to the candidate-conditioned feature-interaction
branch. Specifically, a pyramidal sequence encoder first constructs a
candidate-independent cache from the user's behavioral history.
Candidate-conditioned mix-tokens then repeatedly attend to this cache and
perform Mixup-PerToken-FFN, a simplified variant of the TokenMixer
interaction block. This design assigns clear roles to the two operations: attention
retrieves behavior evidence from a reusable user-side cache, while token-wise
feed-forward interaction updates heterogeneous feature tokens for the current
candidate.

Our contributions are:
\begin{itemize}
  \item We propose \modelname, an interleaved architecture that unifies feature
  interaction and sequence modeling while preserving reusable user-side
  computation, enabling M-FALCON and user-level training amortization in
  production systems.
  \item We provide scaling-law analyses showing that \modelname{} converts
  additional model capacity and compute into consistent ranking-quality gains.
  \item We report significant offline and online improvements from deploying
  \modelname{} in TikTok's e-commerce video recommendation system.
\end{itemize}

\section{\modelname}
\label{sec:method}

Before presenting \modelname, we briefly describe the task setting. We consider
fine ranking in a large-scale e-commerce video recommendation system, where the
ranking model scores a set of candidate videos for a user under the current
context. In production, the model jointly predicts multiple classification and
regression targets, including click, conversion, and GMV-related outcomes. For
a user $u$ and candidate video $c$, the predictions are written as
\begin{equation}
  \hat{\mathbf{y}}_{u,c}
  =f_{\Theta}\!\left(x_c,s_{u},s_{u\times c}\right),
  \label{eq:ranking-task}
\end{equation}
where $x_c$ contains non-sequential user, video, context, and cross features;
$s_u$ denotes candidate-independent U-only behavior sequences;
$s_{u\times c}$ denotes candidate-conditioned U$\times$C behavior sequences; and
$\Theta$ denotes the trainable parameters. The U-only histories are shared by
candidates from the same user request, whereas $x_c$ and $s_{u\times c}$ depend
on the candidate being scored. The task predictions
are subsequently combined into a final ranking score
$s_{u,c}=g(\hat{\mathbf{y}}_{u,c})$, which determines the ordering of candidate
videos, where $g(\cdot)$ denotes the multi-task prediction fusion function.

\subsection{Architecture Overview}
\label{subsec:method-overview}

Given this input partition, \modelname{} maintains three token streams. A
MixTokenizer maps heterogeneous
non-sequential features $x_c$ to $M$ candidate-conditioned mix-tokens. A
SeqTokenizer separately maps candidate-independent user histories
$s_u$ and candidate-conditioned U$\times$C histories
$s_{u\times c}$ to two sequence-token streams:
\begin{equation}
  X_c^0=\mathcal{T}_{m}(x_c),\qquad
  \mathcal{S}_{u}^0=\mathcal{T}_{u}(s_{u}),\qquad
  \mathcal{S}_{u\times c}=\mathcal{T}_{g}(s_{u\times c}),
  \label{eq:token-streams}
\end{equation}
where $X_c^0\in\mathbb{R}^{M\times d_m}$,
$\mathcal{S}_u^0\in\mathbb{R}^{L_u\times d_s}$, and
$\mathcal{S}_{u\times c}\in\mathbb{R}^{L_{u\times c}\times d_s}$. We omit
the batch dimension throughout this section. Here, $M$ is the number of
mix-tokens; $d_m$ and $d_s$ are the hidden dimensions of the non-sequential mix-token and
sequence-token streams; and $L_u$ and $L_{u\times c}$ are the lengths of the
U-only and U$\times$C streams respectively, where U$\times$C denotes sequences
constructed using candidate information, such as hard-search and soft-search
similarity sequences. We further use $N_M$, $N_E$, and $N_D$ to denote the
numbers of MixFormer, SeqFormer Encoder, and SeqFormer Decoder blocks.

Figure~\ref{fig:architecture} shows the resulting computation graph. The
SeqFormer Encoder processes $\mathcal{S}_u^0$ once to construct a
candidate-independent K/V cache. On the candidate side, stacked MixFormer
blocks first alternate retrieval from $\mathcal{S}_{u\times c}$ with feature
interaction over the mix-tokens. Stacked SeqFormer Decoder blocks then
alternate retrieval from the reusable cache with the same feature-interaction
operator. A final mean pooling sends the resulting mix-token representation to
task-specific towers. Crucially, information flows from the user-side sequence
stream to the candidate-side mix-tokens, but never in the reverse direction.
The resulting K/V cache therefore can be shared across all candidates within
the same request during serving and across all samples of the same user during
training, when properly designed.

\subsection{Three-Stream Tokenization}
\label{subsec:tokenization}

\subsubsection{MixTokenizer}
The MixTokenizer converts heterogeneous ranking features into $M$ mix-tokens.
Its inputs include embeddings of all non-sequential categorical and numerical features,
together with a mean-pooled representation of every sequence feature. For each
sequence feature $E_s$, mean pooling along the sequence dimension produces a
fixed-length summary $\bar e_s=\operatorname{MeanPool}(E_s)$. These summaries
provide the mix-token stream with coarse information from all U$\times$C and U-only
sequences, while retaining their token-wise representations for the
SeqTokenizer described next.

The MixTokenizer first applies \bodyemph{All-Concat} to concatenate all
non-sequential feature embeddings and sequence summaries into one vector:
\begin{equation}
  a=\operatorname{Concat}\!\left(
  e_1,\ldots,e_F,\bar e_1,\ldots,\bar e_S\right).
  \label{eq:all-concat}
\end{equation}
It then applies \bodyemph{Auto-Split}, which partitions $a$ into fixed-width groups
and zero-pads an incomplete final group:
\begin{equation}
  [a_1,\ldots,a_M]=\operatorname{AutoSplit}(a).
  \label{eq:auto-split}
\end{equation}
Each group is processed using independently parameterized LayerNorm and
feed-forward projection to produce one mix-token. This per-token
parameterization preserves the distinct semantics and input distributions of
heterogeneous feature groups. Per-token LayerNorm normalizes heterogeneous
token embeddings to a comparable value range, making the tokenizer easier to
optimize. Specifically, we adopt per-token LayerNorm instead of per-token
RMSNorm, as LayerNorm naturally introduces a learnable per-token offset
$\beta$, which plays a role analogous to the field-aware biases in FAT
\cite{yan2025scaling}. All remaining normalization layers use RMSNorm, making
this the only LayerNorm in the entire model.

\subsubsection{SeqTokenizer}
The SeqTokenizer processes two types of sequence features. U$\times$C
sequences, such as similarity-based and item-to-item sequences, incorporate
candidate-side information, whereas U-only sequences contain only user-side
information. During serving with $C$ candidates or training with $J$ samples,
U-only sequence computation can be amortized across candidates or samples,
unlike U$\times$C sequence computation.

Each sequence $s$ is independently mapped to the same sequence dimension $d_s$
by a learned projection matrix $W_s$. Position embeddings are then applied
separately within each sequence:
\begin{equation}
  Q_s=E_sW_s+P_s,
  \label{eq:sequence-tokenizer}
\end{equation}
where $P_s$ is the position embedding specific to sequence $s$. Applying
positions before concatenation avoids imposing a global positional relationship
between elements from different sequences, for which no well-defined relative
order exists across sequence types.

The projected sequences are concatenated according to their dependency type.
Several learnable \texttt{[SINK]} tokens are prepended to the concatenated
streams, marking the beginning of each stream and serving as attention sinks
\cite{xiao2024efficientstreaminglanguagemodels}:
\begin{equation}
  \begin{aligned}
    \mathcal{S}_u^0&=\left[Q_u^{\mathrm{sink}};
    \operatorname{Concat}_{s\in s_u}Q_s\right],\\
    \mathcal{S}_{u\times c}&=\left[Q_{u\times c}^{\mathrm{sink}};
    \operatorname{Concat}_{s\in s_{u\times c}}Q_s\right].
  \end{aligned}
  \label{eq:sequence-streams}
\end{equation}

\subsection{MixFormer}
\label{subsec:mixformer}

As illustrated in the left inset of Figure~\ref{fig:architecture}, each
MixFormer block consists of two consecutive residual sublayers: cross-attention
retrieves candidate-relevant information from U$\times$C sequences, and
Mixup-PerToken-FFN (MPTF) subsequently performs feature interaction among
mix-tokens. An input normalization is applied to the heterogeneous mix-tokens
before they enter the block stack.

MixFormer block $\ell\in\{0,\ldots,N_M-1\}$ computes
\begin{align}
  \widehat X_c^\ell
  &=X_c^\ell+\operatorname{CrossAttn}\!\left(
    \operatorname{RMSNorm}(X_c^\ell),\mathcal{S}_{u\times c}\right),
  \label{eq:mixformer-attn}\\
  X_c^{\ell+1}
  &=\widehat X_c^\ell+
  \operatorname{MPTF}\!\left(
    \operatorname{RMSNorm}(\widehat X_c^\ell)\right),
  \label{eq:mixformer-mptf}
\end{align}
After the final MixFormer block, $X_c^{N_M}$ is passed directly to the SeqFormer Decoder.
We deliberately omit output normalization here, because the following decoder
(Section~\ref{subsubsec:seqformer-decoder}) already normalizes its input.

\subsubsection{Cross-Attention}
All attention modules in \modelname{} use QK normalization, grouped-query
attention (GQA), and gated attention
\cite{qiu2025gatedattentionlargelanguage}. Let $\operatorname{CrossAttn}(X,Y)$ denote the cross-attention sublayer, where $X$ supplies query states and $Y$ supplies key/value
states. For one attention head, we define
gated scaled dot-product attention (SDPA) as follows:
\begin{equation}
  \begin{gathered}
    Q=XW^Q,\qquad [K,V,G]=YW^{KVG},\\
    \bar Q=\operatorname{RMSNorm}(Q),\qquad
    \bar K=\operatorname{RMSNorm}(K),\\
    \widetilde V=V\odot\sigma(G),\\
    \mathcal{A}(Q,K,\widetilde V)=
    \operatorname{softmax}\!\left(
    \frac{\bar Q\bar K^\top}{\sqrt{d_h}}\right)\widetilde V W^O.
  \end{gathered}
  \label{eq:gated-attention}
\end{equation}
where $\sigma$ denotes the sigmoid function.
In MixFormer, $X=X_c^\ell$ contains mix-tokens and
$Y=\mathcal{S}_{u\times c}$ contains U$\times$C sequence tokens. The query
projection matrix $W^Q$ is parameterized per mix-token to preserve
heterogeneity, while U$\times$C sequence tokens use position-wise K/V/G
projections.

\subsubsection{Mixup-PerToken-FFN}
MPTF simplifies the TokenMixer interaction operator
\cite{jiang2026tokenmixer}. Let $X=[x_1;\ldots;x_M]\in\mathbb{R}^{M\times
d_m}$ and assume $d_m$ is divisible by $M$. Split every token into $M$ channel
chunks, $x_j=[x_j^{(1)};\ldots;x_j^{(M)}]$, and define the parameter-free
Mixup transformation as
\begin{equation}
  \widetilde x_t=
  [x_1^{(t)};x_2^{(t)};\ldots;x_M^{(t)}],\qquad 1\le t\le M.
  \label{eq:mixup}
\end{equation}
Each output token therefore receives one channel slice from every input token.
A token-wise FFN with SwiGLU activation then performs nonlinear interaction:
\begin{equation}
  \operatorname{MPTF}_t(X)=
  \left[\operatorname{SiLU}(\widetilde x_tW^g_t)
  \odot(\widetilde x_tW^u_t)\right]W^d_t.
  \label{eq:mptf}
\end{equation}
Mixup provides explicit cross-token communication without the quadratic
attention map used by self-attention, while the token-wise projections retain
feature-group heterogeneity. Repeatedly stacking Cross-Attention and MPTF
enables sequence retrieval and feature interaction in an interleaved style.

\subsection{SeqFormer}
\label{subsec:seqformer}

SeqFormer uses an encoder to transform U-only histories into a reusable user-side
K/V cache. Its decoder enables candidate-side mix-tokens to repeatedly retrieve
information from that cache.

\subsubsection{SeqFormer Encoder}
\label{subsubsec:seqformer-encoder}

As shown in the right inset of Figure~\ref{fig:architecture}, each encoder block
contains two residual sublayers: self-attention followed by a position-wise
FFN.

Before the first block, padded positions are removed once according to the
sequence mask. All subsequent encoder computation operates on the resulting
variable-length representation, avoiding redundant attention and feed-forward
computation on padding. Figure~\ref{fig:rm-padding} illustrates this execution
path, and Section~\ref{sec:optimizations} provides the full implementation details.

Let $\mathcal{S}_u^e\in\mathbb{R}^{L_e\times d_s}$ denote the valid states
entering encoder block $e\in\{0,\ldots,N_E-2\}$, and let $R^e$ denote the retained states used as query and residual branch. The source and retained states are constructed as
\begin{equation}
\begin{aligned}
  \bar{\mathcal{S}}_u^e
  &= \operatorname{RMSNorm}(\mathcal{S}_u^e), \\
  R^e
  &=
  \begin{cases}
    \operatorname{Gather}\!\left(\mathcal{S}_u^e; k_s\right),
    & e = 0, \\
    \operatorname{Truncate}\!\left(\mathcal{S}_u^e, L_{e+1}\right),
    & e > 0,
  \end{cases} \\
  \bar{R}^e
  &=
  \begin{cases}
    \operatorname{Gather}\!\left(\bar{\mathcal{S}}_u^e; k_s\right),
    & e = 0, \\
    \operatorname{Truncate}\!\left(\bar{\mathcal{S}}_u^e, L_{e+1}\right),
    & e > 0.
  \end{cases}
\end{aligned}
\label{eq:seqformer-compression}
\end{equation}
The attention projections are then computed as
\begin{equation}
  \begin{aligned}
    Q^e&=\bar{R}^e W_e^Q, &
    [K^e,V^e,G^e]&=\bar{\mathcal{S}}_u^e W_e^{KVG},\\
    \widetilde V^e&=V^e\odot\sigma(G^e).&&
  \end{aligned}
  \label{eq:seqformer-projections}
\end{equation}
The block update is
\begin{align}
  \widehat{\mathcal{S}}_u^{e+1}
  &=R^e+\mathcal{A}\!\left(Q^e,K^e,\widetilde V^e\right),
  \label{eq:seqformer-encoder-attn}\\
  \mathcal{S}_u^{e+1}
  &=\widehat{\mathcal{S}}_u^{e+1}+
  \operatorname{FFN}\!\left(
    \operatorname{RMSNorm}(\widehat{\mathcal{S}}_u^{e+1})\right).
  \label{eq:seqformer-encoder-ffn}
\end{align}

The encoder further reduces cost by progressively decreasing the number of
query states while retaining all current states as keys and values. It follows
a predefined compression schedule
\begin{equation}
  L_0\ge L_1\ge\cdots\ge L_{N_E-1}=L_c,
  \label{eq:pyramid}
\end{equation}
where $L_c$ is the number of user-side states retained for cache construction.
Gather performs the first compression by extracting recent tokens from every
source sequence. Each sequence receives a token budget $k_s$ determined by
feature-importance analysis, and Gather selects its $k_s$ most recent valid
states before packing all selected states into one query stream. Following the
direct truncation strategy of Ultra HSTU \cite{ding2026bending}, later blocks
retain a shrinking prefix of this stream. The retained query states attend to
all valid states in the current layer before compression.

The final encoder block projects these states into the K/V cache: it omits SDPA, out projection, 
and position-wise FFN, and applies only K/V projection and value gating.
\begin{align}
  [K_u,V_u,G_u]
  &=\mathcal{P}_{KVG}\!\left(
    \operatorname{RMSNorm}(\mathcal{S}_u^{N_E-1})\right),
  \nonumber\\
  \widetilde V_u&=V_u\odot\sigma(G_u).
  \label{eq:seqformer-cache}
\end{align}
The resulting user-side K/V cache is
$\mathcal{C}_u=(M_u,K_u,\widetilde V_u)$. As shown in
Figure~\ref{fig:rm-padding}, the variable-length K/V tensors are padded back to a
static shape with a fixed maximum length, and $M_u$ denotes their valid-position
mask.

Unlike MixFormer and the SeqFormer Decoder, the SeqFormer Encoder uses a shared
position-wise FFN because U-only sequence tokens are comparatively homogeneous.

\subsubsection{SeqFormer Decoder}
\label{subsubsec:seqformer-decoder}

The decoder mirrors MixFormer's two-sublayer structure. It replaces the U$\times$C
sequence input with the reusable user-side cache and first applies an input RMSNorm,
$Z_c^0=\operatorname{RMSNorm}(X_c^{N_M})$. Decoder block
$r\in\{0,\ldots,N_D-1\}$ then retrieves from the cache and applies MPTF:
\begin{align}
  \widehat Z_c^r
  &=Z_c^r+\mathcal{A}_{\mathrm{cache}}\!\left(
    \operatorname{RMSNorm}(Z_c^r),K_u,\widetilde V_u;M_u\right),
  \label{eq:seqformer-decoder-attn}\\
  Z_c^{r+1}
  &=\widehat Z_c^r+
  \operatorname{MPTF}\!\left(
    \operatorname{RMSNorm}(\widehat Z_c^r)\right).
  \label{eq:seqformer-decoder-mptf}
\end{align}
Here, we adopt a lazy K/V strategy: only $Z_c^r$ is projected using the per-token
query projection, while the cached $K_u$ and $\widetilde V_u$ are reused without
additional projection. As $Z_c^r$ is progressively refined, subsequent blocks
can leverage higher-order queries to retrieve richer patterns from the same K/V
cache. Attention is responsible for sequence retrieval, while MPTF is
responsible for feature interaction. Alternating these operations instead of
combining them into a single attention layer prevents mix-tokens from
over-attending to themselves, which could otherwise reduce their ability to
retrieve useful sequential information from the K/V cache. A final RMSNorm
followed by mean pooling produces the representation consumed by the
task-specific towers for prediction.

\section{Experiments}
\label{sec:experiments}

We conduct comprehensive experiments to evaluate \modelname{} on a large-scale
internal industrial dataset. Through offline evaluations and online A/B tests,
we aim to answer the following research questions (RQs):
\begin{itemize}
  \item \textbf{RQ1: Overall effectiveness and efficiency.} How does
  \modelname{} compare with representative production ranking architectures in
  ranking quality and computational cost?
  \item \textbf{RQ2: Scaling behavior.} Does jointly scaling feature
  interaction and sequence modeling produce predictable and sustained quality
  gains as model capacity and compute increase?
  \item \textbf{RQ3: Key design choices.} Which choices in feature interaction,
  numerical stabilization, K/V-cache construction, and attention/residual
  design contribute to ranking quality?
  \item \textbf{RQ4: Production deployment.} Does deploying \modelname{} in a
  live recommendation system yield meaningful improvements in key business
  metrics?
\end{itemize}

\begin{table}[!t]
  \centering
  \caption{Offline comparison on the internal industrial dataset. Quality
  metrics are relative lifts over the RankMixer + Transformer baseline;
  parameters and per-sample training FLOPs are absolute. OneTrans here is evaluated using an updated
  production implementation with post-publication optimizations. In our
  scenario, a relative AUC improvement of 0.03\% is considered
  practically significant and can lead to substantial gains in online metrics.}
  \label{tab:offline-main}
  \small
  \setlength{\tabcolsep}{5pt}
  \begin{tabular}{lrrrrrr}
    \toprule
    & \multicolumn{2}{c}{CTR} & \multicolumn{2}{c}{CVR}
    & \multicolumn{2}{c}{Efficiency} \\
    \cmidrule(lr){2-3}\cmidrule(lr){4-5}\cmidrule(lr){6-7}
    Model & $\Delta\mathrm{AUC}\uparrow$ & $\Delta\mathrm{UAUC}\uparrow$
    & $\Delta\mathrm{AUC}\uparrow$ & $\Delta\mathrm{UAUC}\uparrow$
    & Params & FLOPs \\
    \midrule
    RankMixer + Transformer (baseline) & -- & -- & -- & -- & 210M & 5.75G \\
    \midrule
    RankMixer + Transformer (Large) & +0.04\% & +0.15\% & +0.03\% & +0.19\%
    & 1069M & 27.50G \\
    OneTrans & +0.15\% & +0.61\% & +0.10\% & +0.66\% & 361M & 39.27G \\
    \midrule
    \textbf{\modelname-L} & \textbf{+0.21\%} & \textbf{+0.82\%}
    & \textbf{+0.17\%} & \textbf{+1.38\%} & \textbf{358M}
    & \textbf{35.28G} \\
    \bottomrule
  \end{tabular}
\end{table}

\subsection{Experimental Setup}
\label{subsec:experimental-setup}

\paragraph{Dataset and tasks.}
We conduct offline experiments on an anonymized internal industrial dataset collected from
TikTok's e-commerce video recommendation system. The ranking model jointly
predicts multiple user-response and value targets. Following the production
setting, we report CTR and CVR as representative classification tasks.

\paragraph{Baselines.}
We compare \modelname{} with two baseline architectures in the same industrial
training stack. \bodyemph{RankMixer + Transformer} \cite{zhu2025rankmixer}, evaluated
in standard and large configurations, is the production baseline architecture.
\bodyemph{OneTrans} \cite{zhang2026onetrans} is another strong baseline that combines
feature interaction and sequence retrieval in a unified token stream. We use
an updated production implementation that incorporates post-publication
optimizations. All methods use aligned data, features, prediction tasks, and
optimization protocols.

\paragraph{Metrics.}
Offline effectiveness is measured by AUC and user-level AUC (UAUC). To avoid
disclosing absolute production metrics, we report the relative lift over the
RankMixer + Transformer baseline. Efficiency is characterized by non-embedding
parameter count and per-sample training FLOPs before amortization.
For the online evaluation, we report relative business-metric lifts from a
randomized A/B test against the deployed production baseline.

\subsection{RQ1: Overall Offline Performance}
\label{subsec:main-results}

Table~\ref{tab:offline-main} compares ranking quality and computational cost.
\modelname{} outperforms all compared architectures across the CTR and CVR
metrics. Scaling the RankMixer + Transformer baseline to its large configuration
yields only limited gains, indicating that this two-stage architecture remains
difficult to scale effectively. OneTrans combines both stages in a single
attention stream, but differs from \modelname{} in three respects. First, it uses
self-attention for feature interaction, whereas MPTF is better suited to highly
heterogeneous feature tokens. Second, allowing mix-tokens to attend themselves diverts attention away from the sequence K/V cache, thereby weakening targeted behavior retrieval. Third, OneTrans employs causal attention,
whereas the SeqFormer Encoder uses full, non-causal attention. Causal masking is
unnecessary for discriminative ranking and introduces additional inductive
biases, such as imposing an artificial ordering among input sequences. By interleaving bidirectional cross-attention with MPTF, \modelname{} effectively addresses these limitations.

\subsection{RQ2: Scaling Behavior}
\label{subsec:scaling}

\begin{figure}[H]
  \centering
  \includegraphics[width=0.56\textwidth]{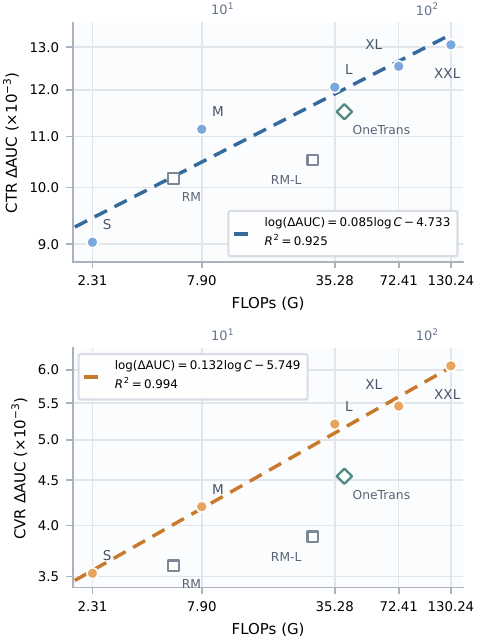}
  \caption{Scaling behavior on CTR and CVR AUC. Dashed lines show linear fits
  between $\log(\Delta\mathrm{AUC})$ and log FLOPs for the observed
  \modelname{} S/M/L/XL/XXL configurations. RankMixer and OneTrans points are
  shown for reference and excluded from the fits. Both axes are logarithmic.}
\Description{Two vertically stacked plots show CTR and CVR AUC gains versus
FLOPs for five HELIX scales and three comparator
configurations. HELIX gains increase with compute, and log-linear fits closely
follow the five HELIX configurations.}
  \label{fig:scaling-law}
\end{figure}

Using aligned data and training settings, we scale \modelname{} across S/M/L/XL/XXL configurations by jointly increasing backbone width, depth, and the retained sequence-token budget. Table~\ref{tab:model-scale} summarizes their
parameter and compute budgets. Figure~\ref{fig:scaling-law} shows that CTR and
CVR AUC improve smoothly with compute and remain unsaturated at the largest
configuration. RankMixer at two scales and OneTrans are included as reference
points at their corresponding compute budgets.
RankMixer remains competitive at low FLOPs but falls behind as compute grows,
while OneTrans trails \modelname-L despite a larger compute budget, indicating
that architectural design determines how effectively additional compute
translates into ranking gains.
The consistent power-law trends indicate predictable ranking-quality gains from
jointly scaling feature interaction and sequence modeling.

\begin{table}[!t]
  \centering
  \caption{\modelname{} model-scale configurations. Quality metrics are
  relative lifts over \modelname-S.}
  \label{tab:model-scale}
  \small
  \setlength{\tabcolsep}{3.5pt}
  \begin{tabular}{lrrrr}
    \toprule
    Model & CTR $\Delta\mathrm{AUC}\uparrow$ & CVR $\Delta\mathrm{AUC}\uparrow$
    & Params & FLOPs \\
    \midrule
    \modelname-S   & -- & -- & 46M   & 2.31G \\
    \modelname-M   & +0.24\% & +0.07\% & 99M   & 7.90G \\
    \textbf{\modelname-L} & \textbf{+0.34\%} & \textbf{+0.18\%}
    & \textbf{358M} & \textbf{35.28G} \\
    \modelname-XL  & +0.39\% & +0.20\% & 740M  & 72.41G \\
    \modelname-XXL & +0.45\% & +0.27\% & 1050M & 130.24G \\
    \bottomrule
  \end{tabular}
\end{table}

\subsection{RQ3: Which Design Choices Matter?}
\label{subsec:ablation}

Table~\ref{tab:ablations} summarizes paired ablations of the design groups that
most directly determine \modelname{}'s effectiveness and stability during
training.

\begin{table}[!t]
  \centering
  \caption{Relative AUC gains from ablations of key design choices in
  \modelname. MPTF is compared against self-attention; K/V Cache-Only Cross-Attention is compared against cross-attention over both mix-tokens and the sequence K/V cache; and the Final-Layer K/V Cache is compared against the Per-Layer K/V Cache. Rows prefixed with $+$ report the gains obtained by adding the corresponding component. Specifically, our final design adopts only gated attention applied after the value projection.}
  \label{tab:ablations}
  \begin{tabular}{@{}>{\raggedright\arraybackslash}p{0.52\columnwidth}rr@{}}
    \toprule
    Design choice & CTR $\Delta$AUC$\uparrow$ & CVR $\Delta$AUC$\uparrow$ \\
    \midrule
    Mixup-\allowbreak PerToken-\allowbreak FFN & +0.05\% & +0.02\% \\
    K/V Cache-only Cross-Attention & +0.02\% & +0.02\% \\
    Final-Layer K/V Cache & +0.01\% & 0.00\% \\
    \midrule
    $+$ Non-parametric Gate-Up-Norm & +0.02\% & +0.01\% \\
    $+$ Gated attention (after value) & +0.03\% & +0.02\% \\
    $+$ Gated attention (after SDPA) & +0.02\% & +0.01\% \\
    $+$ QK-Norm & +0.03\% & +0.01\% \\
    $+$ Grouped-query attention & 0.00\% & 0.00\% \\
    $+$ Encoder LayerScale & +0.01\% & 0.00\% \\
    \bottomrule
  \end{tabular}
\end{table}

\paragraph{Mixup-\allowbreak PerToken-\allowbreak FFN.}
Replacing self-attention-based feature interaction in the SeqFormer Decoder
with MPTF improves CTR and CVR AUC by 0.05\% and
0.02\%, respectively, at a matched parameter and compute budget. Consistent
with TokenMixer \cite{jiang2026tokenmixer}, this result suggests that
self-attention is a poor fit for highly heterogeneous feature tokens: it brings
less quality benefit in this setting while consuming additional compute and
activation memory. We therefore use MPTF for feature interaction.

\paragraph{K/V Cache-only Cross-Attention.}
We remove the mix-token self-attention component from the original cross-attention over $[\text{mix-tokens}, \text{K/V Cache}]$ in OneTrans, restricting mix-tokens to attend exclusively to the K/V cache. This modification yields a 0.02\% AUC gain on both CTR and CVR. We hypothesize that mix-tokens tend to allocate substantial attention mass to themselves, thereby diverting attention away from sequential representations and weakening effective behavior retrieval. By excluding mix-tokens from the key/value set, cross-attention is forced to focus solely on sequence information, enabling more targeted extraction of user behaviors.

\paragraph{Final-Layer K/V Cache.}
Decoder mix-tokens retrieve only from the K/V cache produced by the last layer
of the SeqFormer Encoder. Our experiments show comparable ranking quality to
retrieving from all encoder-layer K/V caches, while storing only the last-layer
cache reduces memory overhead.More importantly, the last-layer K/V cache allows for more flexible computation
allocation between feature interaction and sequence modeling. Per-layer caches
tie each feature-interaction block to a corresponding sequence-modeling layer,
forcing the depths and parameter counts of the two axes to grow together. A
single last-layer cache removes this coupling: sequence modeling can be deepened
without proportionally increasing candidate-side feature interaction, enabling
asymmetric scaling toward the better-amortized sequence branch. The last-layer
design improves CTR AUC by 0.01\% while keeping CVR AUC unchanged.

\paragraph{Non-parametric Gate-Up-Norm.}
The combination of per-token parameterization and SwiGLU's element-wise
multiplication is the primary source of MPTF's training instability. The gate
and up-projection branches can develop heterogeneous activation scales, whose
product further amplifies outliers. We initially use SoftCap to hard-clip
extreme activation values. Although this stabilizes training to some extent,
activations below the clipping threshold can still have large scales and impair
training stability. Inspired by QK-Norm
\cite{henry2020querykeynormalizationtransformers,muennighoff2025olmoe}, we first
introduce \bodyemph{Gate-Up-Norm}, which separately normalizes the gate and
up-projection activations before their element-wise product. However, its
learnable gains can re-amplify the normalized activations, so large activation
values remain. Following the non-parametric normalization used in OLMo
\cite{groeneveld2024olmoacceleratingsciencelanguage}, we remove these learnable
gains to obtain \bodyemph{Non-parametric Gate-Up-Norm}. This design substantially
narrows the activation ranges of the product and down-projection outputs,
keeping them within the FP16 dynamic range and the high-precision region of
BF16 without SoftCap. Non-parametric Gate-Up-Norm stabilizes training and
improves CTR AUC by 0.02\% and CVR AUC by 0.01\%.

\paragraph{Other Refinements.}
Following gated attention \cite{qiu2025gatedattentionlargelanguage}, we compare
two gate placements. Applying the gate after SDPA improves CTR and CVR AUC by
0.02\% and 0.01\%, respectively, whereas applying it after the value projection
improves them by 0.03\% and 0.02\%. We hypothesize that the larger gain from
value-side gating arises because sequence compression leaves the value stream longer than the
query stream, allowing the gate to modulate more sequence states before
attention aggregation.

QK-Norm, which normalizes queries and keys before SDPA, further improves CTR
and CVR AUC by 0.03\% and 0.01\%, respectively.
Grouped-query attention (GQA) \cite{ainslie2023gqa} reduces K/V cache size by
sharing K/V heads across query heads and also lowers the cost of K/V/G
projections. With the same number of query heads, GQA matches multi-head
attention in AUC. Finally, applying LayerScale \cite{touvron2021going} to the
SeqFormer Encoder residual branches improves CTR AUC by 0.01\% with neutral CVR
AUC.

\subsection{RQ4: Online Impact}
\label{subsec:online-results}

We deploy the \modelname-L configuration in TikTok's e-commerce video
recommendation system and evaluate it through a randomized, full-traffic
online A/B test against the RankMixer + Transformer production baseline.

Table~\ref{tab:business-main}
reports relative lifts for the e-commerce video recommendation scenario.
\modelname{} improves video clicks per user by 2.0591\%, video GMV per user by
5.9932\%, video paid orders per user by 6.0964\%, and video GPM by 5.4624\%.
These results confirm substantial production impact from the offline ranking
gains achieved by \modelname{} in deployment.

\begin{table}[!t]
  \centering
  \caption{Online A/B-test lifts for \modelname-L in TikTok's e-commerce video
  recommendation system. RankMixer + Transformer is the production control.}
  \label{tab:business-main}
  \small
  \begin{tabular}{lr}
    \toprule
    Metric & Relative lift$\uparrow$ \\
    \midrule
    E-commerce video clicks / user & +2.0591\% \\
    E-commerce video GMV / user & +5.9932\% \\
    E-commerce video paid orders / user & +6.0964\% \\
    E-commerce video GPM & +5.4624\% \\
    \bottomrule
  \end{tabular}
\Description{Primary online relative lifts for HELIX-L over the RankMixer plus
Transformer production baseline on e-commerce video recommendation metrics.}
\end{table}

\FloatBarrier

\section{Training and Serving Optimizations}
\label{sec:optimizations}

To make \modelname{} efficient at production scale, we optimize two complementary
aspects of its execution. Within the SeqFormer Encoder, pyramidal query
compression reduces the active query length, while padding removal and fused
variable-length kernels eliminate padded computation. In parallel, the
candidate-independent user-side branch enables computation amortization:
M-FALCON \cite{zhai2024actions} shares the K/V cache across candidates during
serving, while user-level training amortizes encoder computation across samples
from the same user during large-scale production training.

\subsection{Pyramidal Query Compression}
\label{subsec:pyramidal-query-compression}

A U-only history contains multiple behavior sequences whose lengths and
predictive importance vary. Processing every valid state as a query in the first
encoder block would make its query-side cost grow with the full history length.
To bound this cost, \modelname{} selects a limited set of query states across these
sequences. Directly truncating the globally concatenated stream would make the
selected tokens depend on sequence order and could over-allocate the budget to
whichever sequence appears first. Instead, we assign each source sequence a
budget $k_s$ determined by feature-importance analysis and use
\bodyemph{Gather} to select its $k_s$ most recent valid states. This choice also
matches the non-uniform information density of behavior histories: recent or
high-intensity actions tend to be more informative.
Pooling methods, such as the uniform pooling used in Funnel-Transformer
\cite{dai2020funneltransformerfilteringsequentialredundancy} or segmented
pooling, perform worse in our setting, likely due to the characteristics of
recommendation histories, where recent behaviors are more informative and
better reflect users' current interests. We then use the selected states as queries and main branch. They attend to all valid
input states as keys and values, allowing the first block to compress the full
history while retaining unselected evidence.

Later blocks utilize \bodyemph{Truncate}, which retains a progressively shorter
prefix at each depth following Equation~\ref{eq:seqformer-compression}. The
retained states serve as queries, while all input states to the current layer
remain available as keys and values; each block can therefore absorb information
from the states it will remove before passing a shorter stream forward. This
produces the schedule in Equation~\ref{eq:pyramid}: early blocks preserve broad
behavior coverage, while deeper blocks concentrate computation on the most
recent states.

\subsection{Variable-Length Sequence Execution}
\label{subsec:variable-length-execution}

\begin{figure}[!t]
  \centering
  \includegraphics[width=\textwidth]{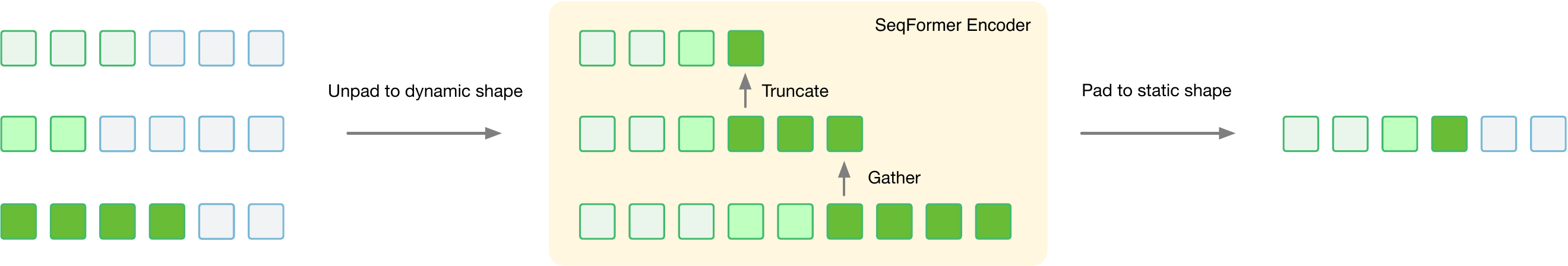}
  \caption{Variable-length execution in the SeqFormer Encoder. Padding is
  removed once; the first compression gathers tokens under per-sequence budgets
  into a single stream, later blocks retain shrinking prefixes, and the resulting
  user-side states are padded back to a static shape for candidate-side
  retrieval.}
  \Description{Padded behavior sequences are packed into a dynamic-shape token
  stream before entering the SeqFormer Encoder. Inside the encoder, gather and
  truncate operations progressively reduce the number of retained tokens. The
  output is padded back to a static shape for candidate-side retrieval.}
  \label{fig:rm-padding}
\end{figure}

As summarized in Figure~\ref{fig:rm-padding}, U-only histories contain substantial
padding because their valid lengths vary across users and sequence types. We
apply the sequence mask once at the SeqFormer Encoder entrance and pack all valid
states. For a batch of $B$ histories with valid lengths
$\{\ell_b\}_{b=1}^{B}$, the encoder operates on $N_v=\sum_b\ell_b$ states rather
than a padded tensor with $B L_{\max}$ positions. This packed representation is
maintained throughout the encoder, so padding is neither materialized nor
processed between blocks.

Data-dependent packed shapes can trigger repeated XLA recompilation, so we
disable XLA for this path and recover efficiency through fused variable-length
kernels for normalization, projections, SwiGLU, residual updates, Gather,
Truncate, etc. The attention sublayers use the variable-length interface of FA4
\cite{zadouri2026flashattention}. Once the encoder constructs its K/V cache, we
pad the cache back to a fixed maximum length and record its valid-position mask.
The SeqFormer Decoder thus receives a static-shape cache while the substantially
more expensive encoder stack remains padding-free.

\subsection{User-Side Computation Amortization}
\label{subsec:user-side-computation-amortization}

Serving and training expose the same reuse pattern. A serving request associates
one user history with multiple candidate items, while training may group multiple
samples from one user with similar histories. Let $C_M$, $C_E$, and
$C_D$ denote the computation costs of MixFormer, SeqFormer Encoder, and SeqFormer Decoder, respectively. Processing $n$ candidate-conditioned instances
independently costs
\begin{equation}
  \operatorname{Cost}_{\mathrm{independent}}(n)
  =n(C_M+C_E+C_D),
\end{equation}
whereas reusing K/V Cache produced by the SeqFormer Encoder costs
\begin{equation}
  \operatorname{Cost}_{\mathrm{reuse}}(n)
  =C_E+n(C_M+C_D).
  \label{eq:reuse-cost}
\end{equation}
The resulting speedup is
\begin{equation}
  \begin{aligned}
    \operatorname{Speedup}(n)
    &=\frac{\operatorname{Cost}_{\mathrm{independent}}(n)}
            {\operatorname{Cost}_{\mathrm{reuse}}(n)} \\
    &=\frac{n(C_M+C_E+C_D)}{C_E+n(C_M+C_D)}.
  \end{aligned}
  \label{eq:reuse-speedup}
\end{equation}
Let $p=C_E/(C_M+C_E+C_D)$ denote the amortizable fraction of the total
per-instance computation, attributable to the SeqFormer Encoder; the remaining fraction is
$1-p=(C_M+C_D)/(C_M+C_E+C_D)$. Equation~\ref{eq:reuse-speedup} can then be
written in the standard Amdahl form:
\begin{equation}
  \operatorname{Speedup}(n)=\frac{1}{(1-p)+p/n}.
  \label{eq:amdahl-speedup}
\end{equation}
The asymmetric modeling discussed in Section~\ref{subsec:ablation} enables a larger fraction of the computation to be allocated to the amortizable SeqFormer Encoder, thereby increasing $p$. When $p=0.95$ and $n=100$, the theoretical speedup can reach approximately $16.8\times$.

\paragraph{Serving with M-FALCON.}
For a request containing $C$ candidates, we set $n=C$ and encode the U-only
history once. The remaining challenge is allowing a microbatch of candidate
queries to access the same cache efficiently. A conventional implementation expands the batch-one cache to the candidate batch size, writes
the materialized copies to high-bandwidth memory (HBM), and reads them again for
attention. M-FALCON \cite{zhai2024actions} instead maps every candidate query
directly to the shared cache inside the attention kernel. It avoids
candidate-wise cache materialization and the associated HBM traffic in the
short-query, long-key regime. This optimization applies to the SeqFormer
Decoder; MixFormer remains candidate-dependent because its U$\times$C sequences
vary across candidates.

\paragraph{Training with user-level training.}
For $J$ samples belonging to the same user, we set $n=J$. User-level training
(ULT) groups these samples and evaluates the U-only encoder once, while their
candidate-conditioned inputs, MixFormer computation, and SeqFormer Decoder
computation remain sample-specific. ULT thus changes the user-side cost from one
encoder evaluation per sample to one evaluation per distinct user, allowing
sequence-modeling capacity to grow without multiplying its full cost by the
number of associated training samples assigned to each user.

\section{Related Work}
\label{sec:related-work}

\paragraph{Scalable recommendation modeling.}
Recommendation ranking models have developed along two complementary scaling
directions. Feature-interaction methods range from classical factorization and
cross networks \cite{rendle2010fm,guo2017deepfm,wang2017dcn} to tokenized
interaction backbones, including Wukong, RankMixer, TokenMixer-Large, and
UniMixer
\cite{zhang2024wukong,zhu2025rankmixer,jiang2026tokenmixer,ha2026unimixer}.
In parallel, sequential recommendation has progressed from target-aware
behavior aggregation \cite{zhou2018din,zhou2019dien,feng2019dsin} to large
transformer-style models such as HSTU, Ultra HSTU, and STCA
\cite{zhai2024actions,ding2026bending,guan2026make}. Together, these lines scale
feature interaction and sequence modeling.

\paragraph{Unified architectures.}
Recent architectures, including InterFormer, Kunlun, OneTrans, SORT, and MTFM
\cite{zeng2025interformer,hou2026kunlun,zhang2026onetrans,wang2026sort,song2026mtfm}, connect
the two modeling axes within a shared or repeated backbone. \modelname{} follows
this joint-scaling direction but gives retrieval and interaction distinct roles:
candidate-conditioned mix-tokens repeatedly retrieve sequence representations
and then undergo feature interaction across depth. Its one-way information flow
keeps user-side sequence computation candidate-independent, so the resulting
K/V cache remains reusable across candidates and samples. This design preserves
practical amortization through M-FALCON and user-level training while allowing
feature interaction and sequence modeling to scale asymmetrically in practice.

\section{Conclusion}
\label{sec:conclusion}

We present \modelname, a purified and unified architecture for large-scale
industrial recommendation that jointly scales feature interaction and sequence
modeling. By interleaving dedicated sequence retrieval with heterogeneous
feature interaction and enforcing one-way information flow, \modelname{} allows
candidate-conditioned mix-tokens to repeatedly benefit from deep user-history
modeling while preserving reusable user-side computation. Its pyramidal
sequence encoder and candidate-independent K/V cache further enable asymmetric
scaling, while padding-free variable-length execution, M-FALCON, and user-level
training make this reuse practical in both training and serving. Furthermore, across the
recommendation cascade, the same cache can be computed once and reused by
generative retrieval, pre-ranking, and ranking, amortizing user-side sequence
computation across stages. Comprehensive offline experiments
demonstrate predictable scaling behavior and validate the key architectural
choices, while deployment in TikTok's e-commerce video recommendation system
confirms that these gains translate into meaningful online business impact.
Together, these results establish \modelname{} as an effective and
production-ready approach to scaling recommendation ranking models along both
modeling axes.

\FloatBarrier
\bibliographystyle{plainnat}
\bibliography{references}

\clearpage
\beginappendix
\section{Contributions}
\label{sec:contributions}

\subsection*{Core Contributors (equal contribution)}
\begin{tabular}{@{}l@{}}
Yuntao Zheng \\
Miao Zhang \\
Yadong Ding \\
Yanchuan Tang \\
Lixiyu Chen \\
\end{tabular}

\subsection*{Contributors}
\begin{tabular}{@{}l@{}}
Hao Wang \\
Quan Li \\
Shiying Cai \\
Yue Lin \\
Jiayu Li \\
Yu Feng \\
Wentao Yang \\
Rongkun Xing \\
Jiekai Wang \\
Mingge Zhang \\
Feiling Gong \\
Xiang Gao \\
Jinyu Dong \\
Yajing Zhang \\
Pengfei Ren \\
Yinzhou Wang 
\end{tabular}

\clearpage
\section{Detailed Online Metrics}
\label{sec:detailed-online-metrics}

Table~\ref{tab:detailed-online-metrics} reports a representative subset of
additional online metrics from the same randomized, full-traffic A/B test as
the primary results in Table~\ref{tab:business-main}. The selected metrics span
business conversion, customer growth and retention, click-through, and
downstream engagement.

\begin{table}[H]
  \centering
  \caption{Representative online A/B-test lifts for \modelname-L over the
  RankMixer + Transformer production control. All values are relative lifts;
  higher is better.}
  \label{tab:detailed-online-metrics}
  \small
  \renewcommand{\arraystretch}{1.08}
    \begin{tabular}{lr}
      \toprule
      Metric & Relative Lift$\uparrow$ \\
      \midrule
    
      \multicolumn{2}{l}{\bfseries Business Conversion} \\
      E-commerce Video GMV per User & +5.9932\% \\
      E-commerce Video Paid Orders per User & +6.0964\% \\
      E-commerce Video GPM & +5.4624\% \\
    
      \addlinespace[2pt]
      \multicolumn{2}{l}{\bfseries Customer Activity and Growth} \\
      E-commerce Purchase Days & +4.3789\% \\
      E-commerce New Paying User Rate & +5.8657\% \\
    
      \addlinespace[2pt]
      \multicolumn{2}{l}{\bfseries Repurchase} \\
      Repurchase Days per Paying User & +0.6617\% \\
      7-Day Repurchase Rate (RPR) & +0.6927\% \\
      30-Day Repurchase Rate (RPR) & +0.6661\% \\
    
      \addlinespace[2pt]
      \multicolumn{2}{l}{\bfseries Click-Through} \\
      Video Anchor Click-Through Rate & +1.8174\% \\
    
      \addlinespace[2pt]
      \multicolumn{2}{l}{\bfseries Engagement} \\
      E-commerce Video Watch Time & +0.9931\% \\
      E-commerce Video Product Detail Page Dwell Time & +2.1324\% \\
      E-commerce Video Like Rate & +0.8489\% \\
      E-commerce Video Follow Rate & +2.5940\% \\
      E-commerce Video Share Rate & +1.7905\% \\
      E-commerce Video Favorite Rate & +1.6013\% \\
    
      \bottomrule
    \end{tabular}
\end{table}

\end{document}